# Cognitive factors that affect the adoption of autonomous agriculture

S. Kate Devitt[1]

[1] Queensland University of Technology

## Abstract

Robotic and Autonomous Agricultural Technologies (RAAT) are increasingly available yet may fail to be adopted. This paper focusses specifically on cognitive factors that affect adoption including: inability to generate trust, loss of farming knowledge and reduced social cognition. It is recommended that agriculture develops its own framework for the performance and safety of RAAT drawing on human factors research in aerospace engineering including human inputs (individual variance in knowledge, skills, abilities, preferences, needs and traits), trust, situational awareness and cognitive load. The kinds of cognitive impacts depend on the RAATs level of autonomy, ie whether it has automatic, partial autonomy and autonomous functionality and stage of adoption, ie adoption, initial use or post-adoptive use. The more autonomous a system is, the less a human needs to know to operate it and the less the cognitive load, but it also means farmers have less situational awareness about on farm activities that in turn may affect strategic decision-making about their enterprise. Some cognitive factors may be hidden when RAAT is first adopted but play a greater role during prolonged or intense post-adoptive use. Systems with partial autonomy need intuitive user interfaces, engaging system information, and clear signaling to be trusted with low level tasks; and to compliment and augment high order decision-making on farm

# Table of Contents

# Introduction

Agricultural Technologies (AgTech) are increasingly digital, smart and automated. Smart farming combines sensors, Internet of Things (IoT), Artificial Intelligence (AI) and robotics to save labour, reduce inputs, and improve livestock wellbeing (Gardenier et al., 2018). Some systems are already independent of human oversight at lower levels of decision making, e.g. auto-steering through paddocks, auto-seeding, selective weed picking (McCool et al., 2018; Milioto et al., 2018; Sa et al., 2018), yield estimation (Riggio et al., 2018), irrigation (Thayer et al., 2018) and harvesting (Mutschler, 2018), and manufacturers are aiming to become more independent at higher decision levels. Robotic and Autonomous Agricultural Technologies (RAAT) such as driverless tractors, fruit picking robots, UAVs (Drones) are increasingly available yet may fail to be adopted (Aubert et al., 2012; Tey & Brindal, 2012). Many reasons explain why RAAT might *theoretically* benefit the farming enterprise yet fail to be *pragmatic* such as cost, performance, mobility and effort required to calibrate, use and maintain technologies. Some limits to higher levels of autonomy are broader infrastructure needs such as high-speed internet and interoperability. This paper focusses specifically on *cognitive* factors that may affect RAAT adoption including:

1. inability to generate trust
2. loss of farming knowledge and
3. reduced social cognition.

The challenge and the opportunity of AgTech is to make RAAT that benefit users at a cognitive level. Such as *reducing* the amount of low level, procedural decision making and *enhancing* higher level, strategic and reflective decision making. Systems that take too much work or require farmers to radically change their management style and preferences face particular psychological barriers.

This paper will first define levels of autonomy and stages of adoption. It will then examine factors that affect the adoption of technology in agriculture more broadly and focus in on particular issues for RAAT and the cognitive factors that affect adoption. The paper finishes with a performance and safety model to assist engineers and policy makers to design RAAT for adoption with cognitive factors in mind.

# Background

## Levels of Autonomy

RAAT are not a single kind of technology because they vary on how independently they operate from human control. There are currently no agricultural technologies that require no human intervention at all—that is are truly 'autonomous'. But, there are increasingly RAATs that operate and control many of their own operations. The less a human is involved, the more RAATs must be capable of traditionally cognitive tasks, that is, to perceive their environment, make decisions, modify behaviour and change goals adaptively (Clothier et al., 2013). There are three types of autonomy that need to be understood: *automatic*, *autonomy* and *autonomous* (See Concepts for UAS Autonomy in Clothier et al., 2013).

## Automatic

*Automatic* or *automated* technologies are able to execute pre-defined processes or events such as auto-steer in a tractor or completing a specific irrigation program. Automation requires farmer initiation, monitoring and/or intervention. An automatic system receives input from sensors and follows rules to obtain a programmed outcome. These rules make the behaviour of systems predictable. Automatic systems may manage some variation in operation, but are not able to define their own goals or to pick their own path based on being given a goal. (Clothier et al., 2013)

## Partial Autonomy

Technologies with *partial autonomy* are able to perceive their environment, manage uncertainty, reason about courses of action and make decisions that lead to unscripted actions (Clothier et al., 2013). These technologies work with humans, where humans set the parameters and goals of operation through a human-robot interface (HRI). Almost all of the robots currently considered 'autonomous' are in fact limited in their autonomy to meet human goals. However, partially autonomous systems can be complex and capable of thinking and reasoning in their environments of use. In agriculture, tractors that are able to map and pick their own paths over a field, detect and remove weeds (manually, chemically or with microwaves), avoid obstacles, re-route to alternate paths and go back to their recharging station when their batteries are low—have autonomy (Bawden et al., 2017).

## Autonomous

A truly autonomous system is capable of higher order cognitive processes including higher level intent and direction. It does not require human oversight and control. Remote farming, where farmers are not physically present or virtually directing the activities on farm would classify as autonomous systems. In this paper I will only consider systems at the current technology readiness of *partial autonomy*. That is, human-robot technologies where the robot has some level of decision making and control, yet is still directed and constrained by human goals and oversight.

## Phases of adoption

Adoption of RAAT will occur in three different phases, each with potentially different barriers:

1. User adoption
2. Initial use
3. Post-adoptive use

### User adoption

User adoption is the initial stage of technology acceptance; when consumers form the intention to use a technology and act on that intentions including purchasing and installation. Moving from non-use to use requires users to meet a threshold of risk, beliefs, social acceptance and emotions that leads them to judge a likely overall advantage over disadvantage. Less risk adverse users may have some level of excitement or willingness to take the gamble on the technology to find out. The greatest barriers to adoption at this stage include the cost of technologies and lack of belief that the technologies will bring an overall advantage.

### Initial use

Initial use involves set up activities and early forays into using the technology and witnessing its limits including:

- *Calibration* to individual farm parameters (e.g. GPS coordinates, soil, water, crop type, seed type, crop reflectance etc...)
- *Personalization* to individual farmer needs and preferences (e.g. speed, accuracy, level of detail)
- *Learning* how to use the technology (e.g. experimentation, workshops, product manuals and/or personal assistance)

Users in the initial stage are often willing to put in a lot of energy to try and make the technology work. The greatest barriers to adoption at this stage include poor user interface design that make set up laborious, complex or confusing; obtuse instructions, or farmer inability to get technology operational. Robotics engineers need to work side-by-side with end users throughout the design phase to improve adoption at the initial use stage. Designing *with*, rather than designing *for*. At the 'initial use' stage users may stop using features or attempt to return or sell a technology. For example, a farmer might buy a drone with a clear sense that the product will enhance on-farm surveillance, yet they find the app on their phone clunky to use, the battery doesn't last long enough to surveil their whole property, they don't have enough storage to manage images, or they find controlling it more difficult and unintuitive than they expected.

## Post-adoptive use

Post-adoptive use is when AgTech is incorporated into the regular work flow of users. It begins after the initial flurry of activity subsides. Users that continue to use a technology into a longer term have found a genuine usefulness in the technology within their needs. Barriers to adoption at this stage includes the failure of technology to adapt to changing user needs and poor support services to help users use the product or to maintain the technology when it needs servicing or repairs. Users may find information management burdensome leading to poor use, such as not downloading or analysing farm surveillance data from a drone. When one aspect of a system is too time-consuming to use, it may stop use of the system overall. E.g. farmers may stop using their drone once they realise they do not have time to analyse or make decisions from the data it is gathering.

# Factors affecting the adoption of technology in agriculture

In this paper we recommend policy-makers refer to the *Unified Theory of Acceptance and Use of Technology (UTAUT)* (Venkatesh et al., 2003; Venkatesh et al., 2016) to frame expectations around the adoption of AgTech—see *Figure 1.* UTAUT can guide the design and regulation of new agricultural technologies. The main factors that affect the adoption of technology are expressed in *Table 1.* Note we have not gone into details regarding gender, age, experience or voluntariness of use, but note that younger, more highly educated and more tech savvy farmers with larger farms are more likely to adopt agricultural technologies (Larson et al., 2008).

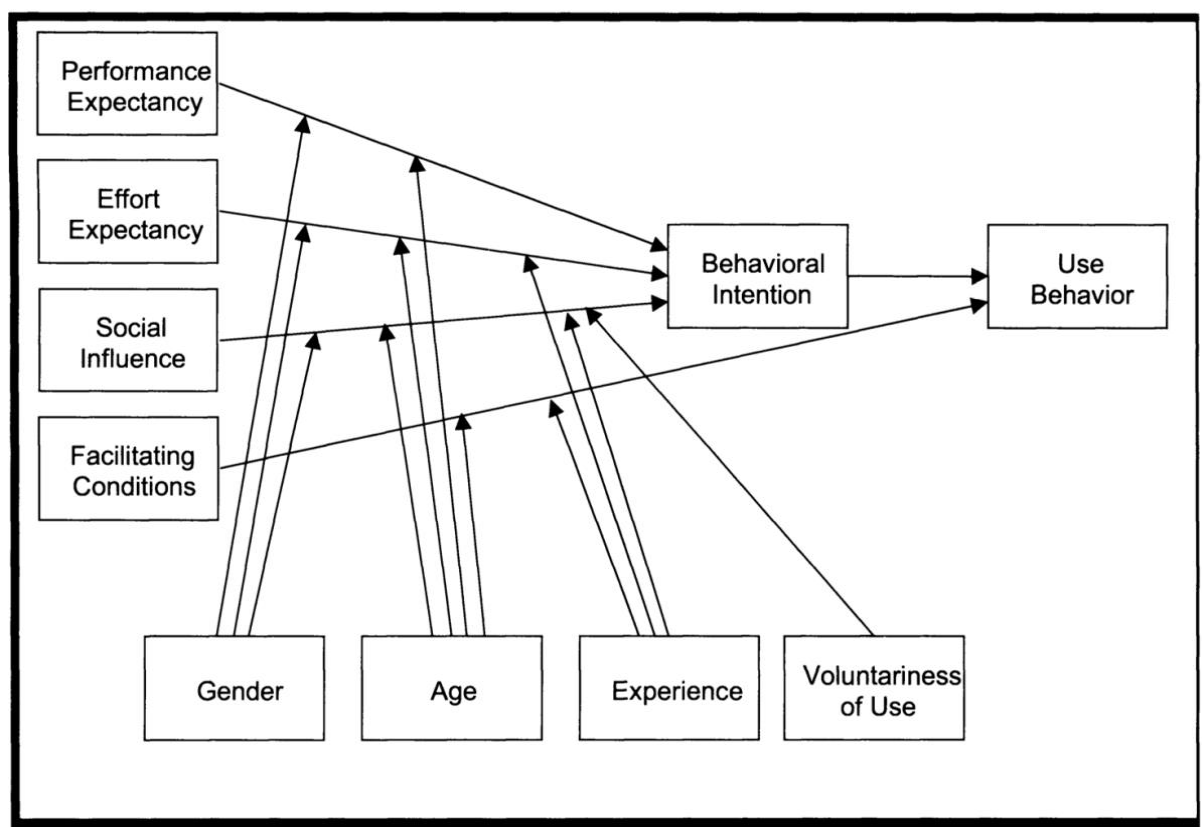

*Figure 1:* Unified Theory of Acceptance and Use (Venkatesh et al., 2003)

While UTAUT is a helpful and robust framework within urban organisational work contexts, it is worth recognising that traditional agricultural decision making is typically made to avoid risk, rather than optimise efficiency (Tey & Brindal, 2012). Farm relevant cognitive factors such as risk aversion are not explicitly represented in UTAUT as potential barriers to adoption. Additionally, robots and autonomous systems have particular factors that make them at risk for adoption. Both of these aspects will now be addressed.

*Table 1:* Main factors that affect the adoption of technology

| | |
|---|---|
| **Performance expectancy** | *Perceived usefulness* of AgTech on farm, i.e. increases in efficiency and effectiveness<br><br>*Extrinsic motivation* such as saving time, benefits to yield, profits and sustainability<br><br>*Relative advantage* to enterprise from adopting technology from existing AgTech<br><br>*Outcome expectations* from using AgTech on the quality and quantity of outcomes |
| **Effort Expectancy** | *Perceived ease of use* is the degree to which a user believes a system will be free of effort or difficult to use<br><br>*Complexity* is the degree to which a system is perceived as relatively difficult to understand and use |
| **Social Influence** | *Subjective norm* relates to a farmer's perception that most of the people who are important to them think they should or should not use the system<br><br>*Social factors* include farmers internalisation of their industry's subjective culture (e.g. protecting jobs), and specific interpersonal agreements and expectations that farmers have made with others in specific social situations, e.g. with agronomist.<br><br>*Image* refers to whether using a technology enhances or detracts from farmer's reputation or status. |
| **Facilitating conditions** | *Perceived behavioural control* is whether farmers believe they have control over the technology including sufficient knowledge to manage it, access to resources to fill knowledge gaps and the system has interoperability with existing systems managed by the farmer<br><br>*Facilitating conditions* is the level of personalised support available to transition the farmer into using the technology, such as the quality of machinery and digital supplier relationships and services and the technology readiness of consultants and extension officers<br><br>*Compatibility* means that technology is consistent with existing values, needs and experiences of farmers |

# Factors affecting the adoption of autonomous systems in agriculture

Robots and autonomous systems are not like other kinds of AgTech. Robots are often brittle and require human supervision and intervention to manage unexpected situations (Wang & Zhang, 2017). Humans out-compete RAATS at high-level decision-making they offer a significant advantage in performing cognitive tasks in complex, dynamic, and uncertain environments (Wang & Zhang, 2017). The highest level of autonomy typically achieved by RAAT is where humans set the hardware and algorithm parameters and the robot is able to perform *limited* tasks without human intervention (Bac et al., 2014). Even so, robots are excellent at processing large volumes of data and sustaining attention for long periods of time. Humans find repetitive tasks tedious and burdensome and are inefficient. Repetitive tasks adversely affect worker's health and safety that leads to high turnover of agricultural employees and motivating interest in RAATs.

Autonomous systems in agriculture have some benefits for adoption over urban autonomous systems. For example, self-driving tractors operating in sparse rural environments face less regulatory pressures than those used on private farming land. Also, agricultural robots would be well-placed to take over low skilled physical work for long hours in remote locations that humans often prefer not to do (The National Farming Union, 2017). However, Agricultural environments are also complex and highly variable posing particular problems for engineers such as: damage from weather, mud and livestock, occlusion of sensors by dirt, unpredictable forms of plants, and variable light conditions (Baxter et al., 2018). RAATS still perform at below-human levels of seeing and grasping and they struggle to perform in dynamic and unstructured environments due to inherent uncertainties, unknown operational settings and unpredictability of events and environmental conditions (Bechar & Vigneault, 2017). Add to this that agricultural produce is highly sensitive to environmental and physical conditions such as temperature, humidity, gas, pressure, abrasion and acceleration. Therefore most fruit, vegetable and flower growing and similar production tasks e.g. trellising, harvesting, sorting and packaging are still performed manually (Bechar & Vigneault, 2017).

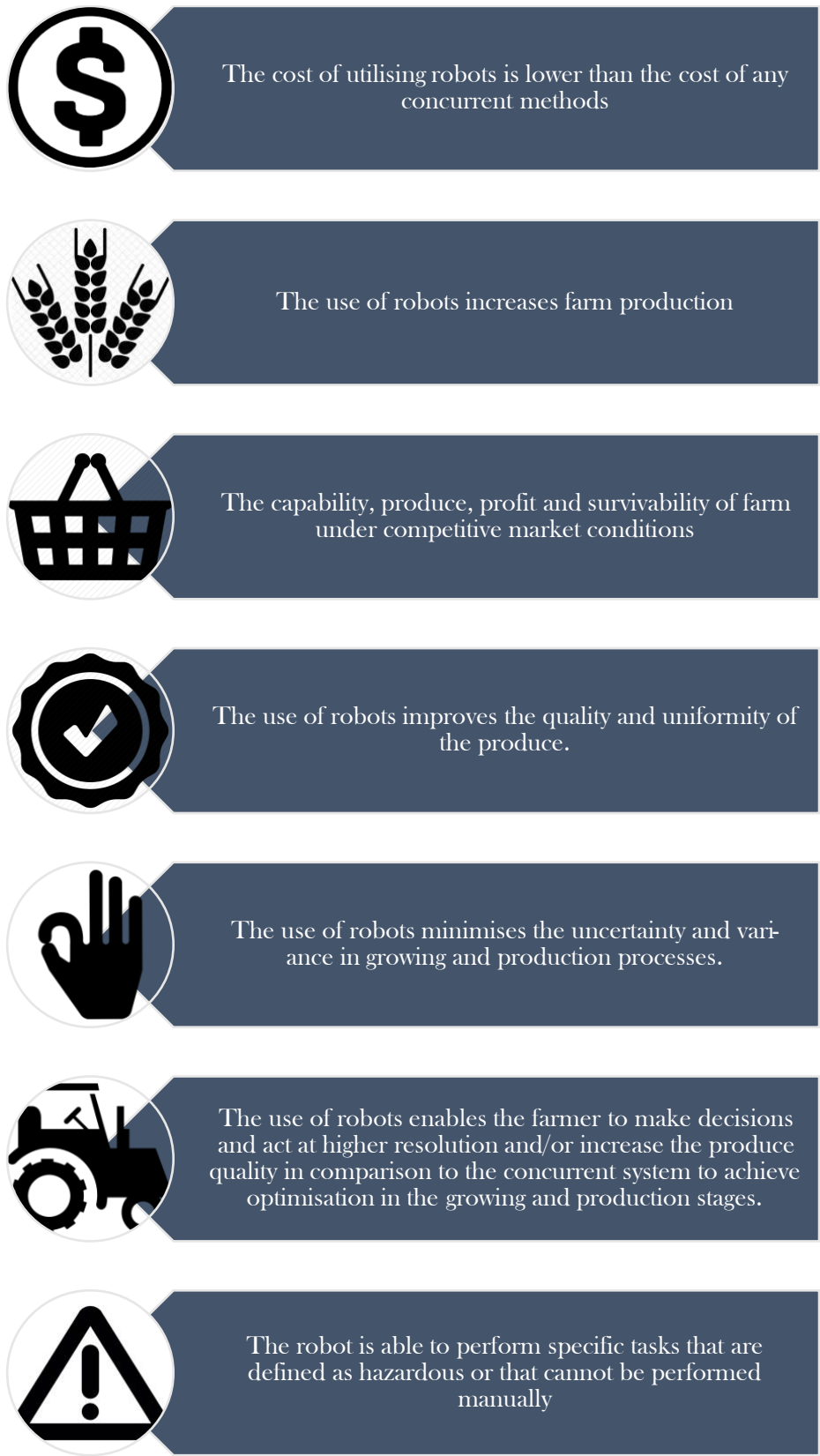

*Figure 2:* Conditions required for the implementation of robotics technology in agriculture (if at least one obtains) (Bechar & Vigneault, 2016)

Finally, when co-located and working with humans additional safety and efficiency requirements must to be considered. (Baxter et al., 2018). Researchers have found that agricultural robotics and intelligent automation are often not implemented because of cost, inability to execute specified agricultural task, low system durability, and inability to successfully reproduce tasks in slightly different contexts or to satisfy mechanical, economic and/or industrial aspects. However, RAATs may be adopted if at least one of the conditions in *Figure 2* are met.

# Cognitive Factors affecting the adoption of autonomous systems in agriculture

Cognitive barriers to the adoption of RAATs include inability to generate and maintain trust with users, loss of farming knowledge and lost social cognition. Note, the relevance of cognitive variables to adoption may only become known during prolonged or intense systems interaction (Gao & Lee, 2006).

## Inability to generate trust

Trust is a complex construct that affects how humans relate to one another and whether they are willing to offload tasks to another agent—whether that agent is human or artificial. I argue that for RAAT to gain trust they need to meet two components: *competency* (comprising of skills, reliability and experience) and *integrity* (comprising motives, honesty and character)—see *Figure 3*. Integrity for RAAT arises from the intentions and behaviour of the design team liable for the production and implementation of the system in situ. RAAT manufacturers need to align products with their corporate or government codes of ethics and make these transparent for end users to ensure their reputation.

What is interesting about the two-component model is that there is an asymmetry between violations of competence versus integrity. If an agent performs incompetently in a task, say a robotic fruit picker gets stuck navigating to the orchard due to a fallen branch or muddy path, it may be frustrating for the human operator to manage, but may not destroy trust. Instead, the human in charge of the robot will take the mistake into consideration and change their work processes to accommodate for it, e.g. conducting a manual sweep of the robot's picking area before setting it off, picking up debris, and putting temporary bridges over problematic pathways, etc… So long as the robot maintains *integrity*, the human is likely to try and work with it to improve outcomes or to continue to buy upgraded models that overcome prior limitations. However, if the robot acts mysteriously or has misaligned values (e.g. collected data is protected by the corporation and unavailable to the farmer without additional payments), the human may no longer trust it—a loss of integrity. Mysterious operations might arise from machine learning algorithms that direct the robot to act in ways confounding to the human operator. The human might wonder, "why is this robot picking fruit in x, y and z ways, when best practice suggests it should pick fruit p, q and r"? If a human does not understand why the agent is operating in particular ways—particularly if it acts contrary to the way a human would act—it can have a profound impact on trust that leads to rejection of the technology.

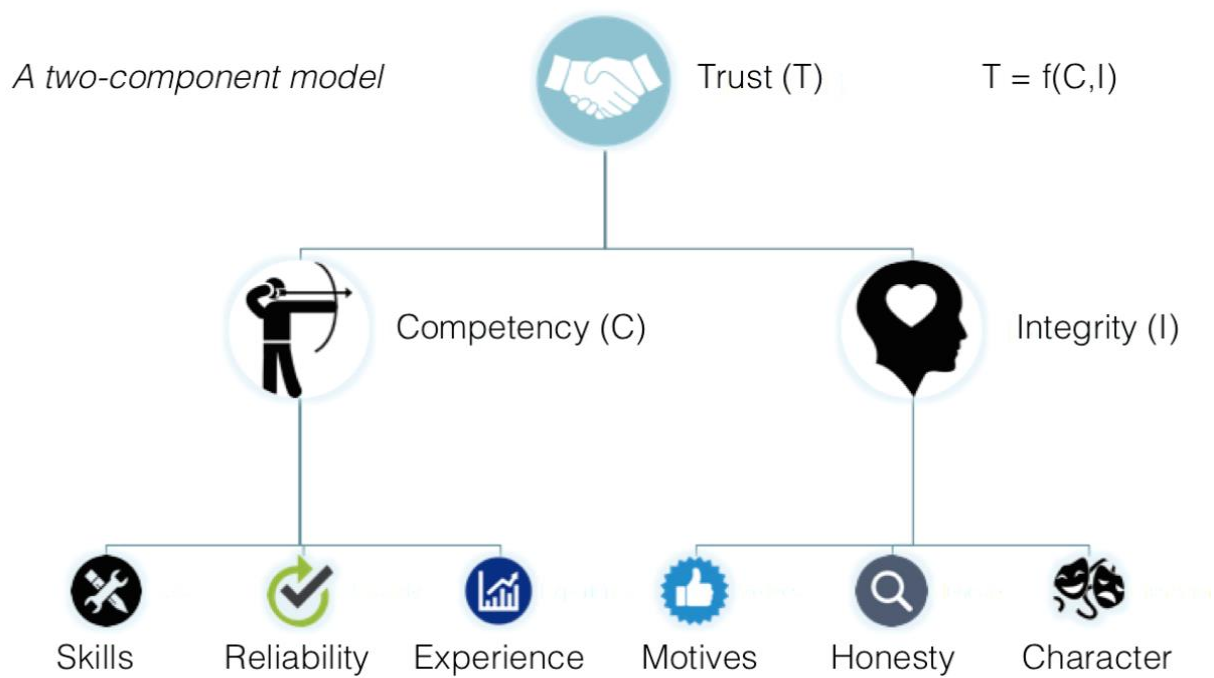

*Figure 3*: A two-component model of trust T = f(C,I). Where Trust comprises of competency (skills, reliability and experience) and integrity (motives, honesty and character) (Devitt, 2018)

## Loss of farming knowledge

The more autonomous agricultural systems become, the less humans may be required on their land to conduct operations. In some circumstances, this may be a huge bonus for farmers. For example, automatic irrigation monitors free up growers from manual checks of their paddocks so that they can save time and labour. Growers freed from certain manual tasks may have more time for strategic thinking, complex problem-solving and optimisation practices on farm such as integrated pest management, nutrition or marketing. However, high levels of automation bring cognitive risks to enterprise perception, knowledge and understanding. Farmers glean much information about their farms when they're physically on-farm, whether in a tractor, checking irrigation channels, manually inspecting a crop or livestock or monitoring equipment. In addition to the specific task at hand, farmers mentally work in parallel, using multiple modalities (e.g. sight, sound, smell, proprioception, touch) to gather information about their farm, prepare scenarios and strategies to run through at a later time with their family, business partner or agronomist. For example, farmers cross-correlate their own intuitions about weather with on farm weather stations, which allows them to be skeptical—where appropriate—and to evaluate risks. If farmers do not directly perceive their own farm and create sophisticated knowledge representations, then they may struggle to imagine or plan possible future actions on-farm in a meaningful way—it depends on the limits of data collected by RAAT and their user interface (UI). One of the dangers of engineer-led design is that data collected doesn't resonate with farmers. For example, growers might trust UAVs to map their fields, but then not manage to download the data collected by them, or they download data, but make little or no time to access it to inform decision-making. Even if farmers do interrogate the data, they may not relate to it psychologically. The way RAAT communicate their actions and findings affects the cognitive processes of farmers. Simplistic reports may fail to trigger complex and informative imaginings and memories necessary to calibrate and generate sophisticated scenarios, actions and strategic decisions.

More and more sensing technologies are becoming available to farmers that could be used as information inputs to guide the actions of RAATs. Take the example of GPS and motion tracking of dairy cattle on farm (Draganova et al., 2010; Goense, 2017; Jukan et al., 2017). Wearables are attached to cows to help farmers monitor activity such as resting behaviours so critical to dairy herd physical and mental wellbeing. Information from trackers can help farmers manage standoff pads and assist cows in poor health sooner and more responsively, e.g. autonomous shepherds and gates could move cows to more comfortable areas if they are under stress and not resting enough. However, these systems are limited in what information they can provide., e.g. they cannot convey distress experienced by cows expressed acoustically rather than physically. Thus, using sensor and RAATs means appreciating the limits of the information they convey and augmenting it with targeted on-the-ground observation and care. When the limits and capabilities of both RAATs and human operators is understood and managed, then machine-human management can be optimised.

## Social cognition

Farming cognition and decision making often occurs as a shared process between key stakeholders on farm such as between growers, agronomists, business partners, and family.

One of the risks of RAATs on farm is lost social cognition. Consider what makes a good agronomist-farmer working relationship: Of course they share information; the agronomist informed in agronomic principles, scientific developments and experiences across multiple farms across a region; the farmer with their background knowledge, values, history of farming decisions, and strategic plans. They know about themselves—what they know and what they don't know—and they each know about each other—what they know and don't know. There's a kind-hearted banter in a good working relationship that allows subtle probability assessment changes for the consequences of decisions. Now, consider autonomous agriculture as it progresses to greater levels of AI. Suppose some (or all) of an agronomist's functions become programmed into an AI available on your smart phone. This AI has access to your farms sensors, tractors, and log data through IoT. It has access to agronomic knowledge and the latest scientific best practice. But, how does the AI impact the *cognition* of the farmer? That is to say, does the farmer spend as much time with the AI asking questions, being skeptical, wondering about this and about that? Does the farmer trust the AI with his or her fears and longings? Does the AI care about the farmer as a person? If AIs don't care and farmers don't really trust them, then there is a great risk with replacing humans with artificial systems. Social cognition is more than an exchange of information, it is a shared system of care and consideration critical for higher order cognitive tasks such as reflection, imagining, motivation, planning and strategic decision-making. Until RAATs are able to have their own goals, they ought to be considered simply in terms of human-robot functionality to extend human goals. Instrumental human-robot relationships increase the fit between person and environment—affordances (Gibson, 1979). One way then to characterize RAATs is self-regulatory technologies to increase affordances in a variety of contexts (Koole & Veenstra, 2015; Shalev & Oron-Gilad, 2015). In other words, the agronomist and farmer dyad should be maintained, even if the nature of their work changes considerably with the development of autonomous agriculture.

# Designing autonomous systems for agriculture

The design and implementation of autonomous systems for agriculture needs a framework to anticipate and scope performance and safety. A model from aerospace engineering human factors research—see *Figure 4*—is recommended as a starting point because these frameworks are designed for remote conditions (e.g. space operations) under high risk. One of the reasons farmers may not adopt technologies is due to the perceived risk of adopting them on their enterprise, e.g. the costs of failure of an autonomous irrigation system for a crop grower includes yield, profits and potentially also reputation. Additionally, farms are frequently remote from services and assistance meaning that the operation of RAAT must be rigorously tested in a large range of conditions.

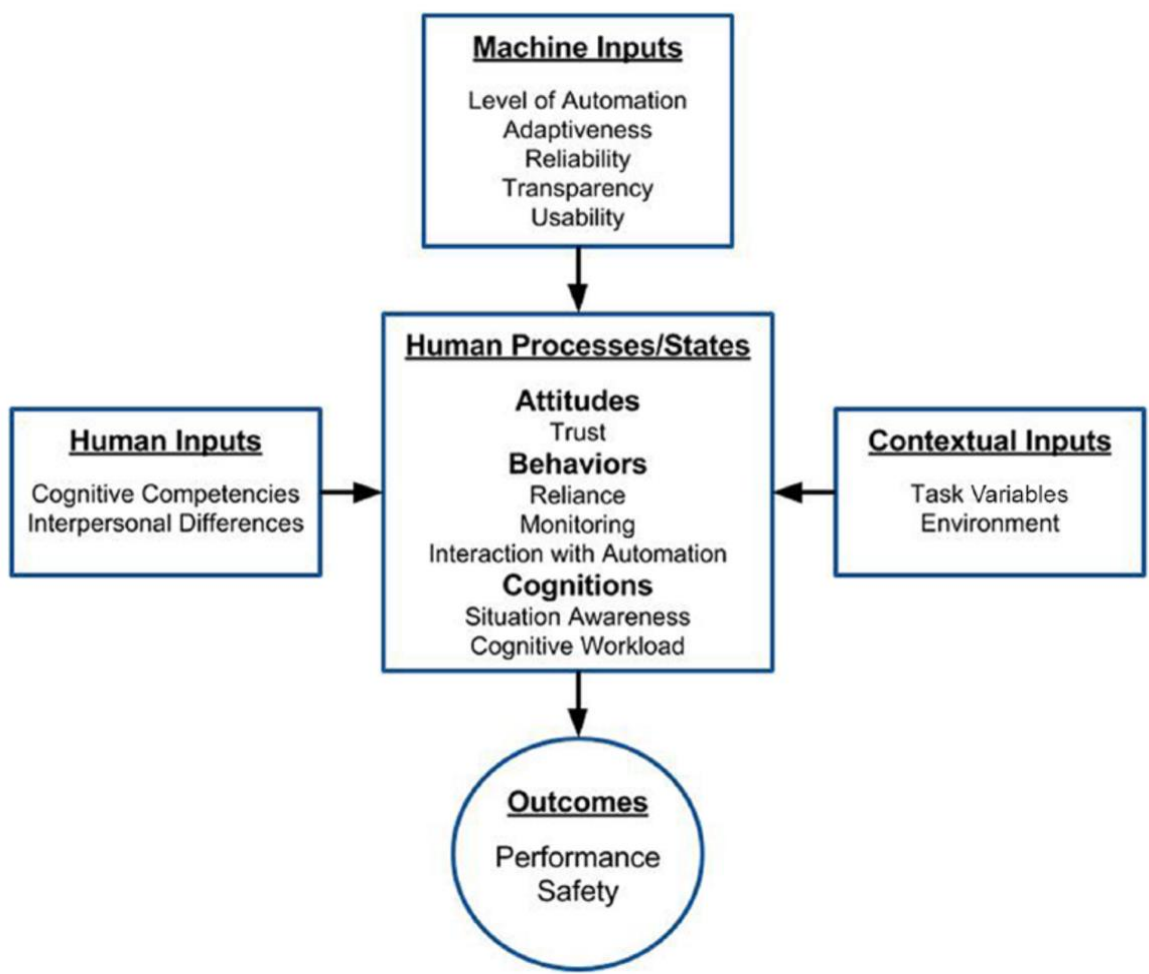

*Figure 4.* Framework of factors affecting safety and performance in human–machine systems (Stowers et al., 2017).

System designers, users, and stakeholders need to use agreed criteria to assess the safety and performance of RAAT, but also consider and assess factors *influencing* safety and performance (Stowers et al., 2017).

*Table 2.* Cognitive Factors affecting safety and performance in human–machine systems (Stowers et al., 2017).

| Criteria | Description |
|---|---|
| Human inputs<br><br>cognitive competencies, interpersonal differences | *Experience and knowledge*, e.g.<br>    agronomic, mechanical, finance, IT, systems<br>*Skills and abilities* e.g.<br>    short-term and long-term memory, probabilistic and abductive reasoning, spatial ability) and<br>*Traits and preferences*<br>    risk aversion across different domains, propensity to trust, desire for control and autonomy vs dependence, preferred social interaction, entrepreneurial vs. managed, contextual impacts on cognition, |
| Attitudes—trust | Trust exists when a human believes a machine will help achieve his or her task or goal. Trust calibration—i.e. learning the *appropriate* amount of trust—leads to appropriate reliance on autonomous systems enhancing safety and performance. Trust calibration can be established through training and exposure to machines. |
| Situational awareness | Performance and safety are at risk when humans lack perception, understanding or ability to predict of their surroundings. The greater the autonomy of a system, the more threat to situational awareness. That is, humans-in-the-loop may lack control, appropriate skill or full awareness of the machine's operations to manage their situation (known as skill decay). |
| Cognitive workload | The relationship between cognitive resources required for a task versus the cognitive resources available to the operator to complete the task. If workload is too high, operators can either cease using the equipment or adapt the task to manage workload. If workload is too low, operators can become bored, distracted and fail to respond effectively to changing events—a particular risk as autonomy of technologies increases. |

This paper recommends policy-makers make sure they understand the cognitive criteria to enhance adoption of autonomous agriculture including human inputs, attitudes, situational awareness and cognitive overload—see *Table 2*. It is recommended that agriculture develops its own framework for the performance and safety of RAAT that incorporates cognitive factors.

# Conclusion

This paper has examined the cognitive factors that may lead to the lack of adoption of robotic and autonomous systems in agriculture. Some systems may see early stage adoption based on reasonable belief that they will improve outcomes on-farm in alignment with farmer values. However, unless systems consider cognitive impacts including lost trust, loss of farmer knowledge and lost social cognition, they may fail to reach post-adoption phase of acceptance and use. Some cognitive factors may be hidden when RAAT is first adopted but play a greater role during prolonged or intense post-adoptive use. The kinds of cognitive impacts depend on the RAATs level of autonomy. The more autonomous a system is, the less a human operator needs to know to operate it and the less the cognitive load, but it also means farmers have less situational awareness about on farm activities that in turn may affect strategic decision making about their enterprise. Systems with partial autonomy need intuitive user interfaces, engaging system information, and clear signaling to compliment and augment high order thinking and decision making for optimal on farm results. Engineers and policy makers must co-design RAATs with end-users, in particular anticipating adoption issues arising from ignoring cognitive factors.